\documentclass[prl,superscriptaddress,showpacs,twocolumn]{revtex4}
\usepackage{graphicx}
\usepackage{amsmath}

\def\vo2{VO$_2$}
\def\vcro2{V$_{1-x}$Cr$_x$O$_2$}

\def\a1g{$a_{1g}$}
\def\t2g{$t_{2g}$}

\input{tcilatex}

\begin{document}

\title{ Spin-dependent Hedin's equations }

\begin{abstract}
Hedin's equations for the electron self-energy and the vertex were
originally derived for a many-electron system with Coulomb interaction. In
recent years it has been increasingly recognized that spin interactions can
play a major role in determining physical properties of systems such as
nanoscale magnets or of interfaces and surfaces. %% or other examples???
We derive a generalized set of Hedin's equations for quantum many-body
systems containing spin interactions, e.g.~spin-orbit and spin-spin
interactions. 
The corresponding spin-dependent GW approximation is constructed.
\end{abstract}

\author{F. Aryasetiawan}
\affiliation{Research Insitute for Computational Sciences, AIST, 1-1-1 Umezono, Tsukuba
Central 2, Ibaraki 305-8568, Japan}
\author{S. Biermann}
\affiliation{Centre de Physique Th{\'e}orique, Ecole Polytechnique, CNRS, 91128 Palaiseau
Cedex, France}
\maketitle

%\pacs{71.27.+a, 71.30.+h, 71.15.Ap}

An increasing number of modern applications of materials science imply spin
degrees of freedom. Spintronics devices, colossal magneto resistance
materials or magnetic impurities in semiconductors are important examples,
not only due to their technological interest but also owing to the
fundamental questions they raise. In fact, a detailed understanding of such
systems requires a theoretical description of the interplay of spin, charge
and orbital degrees of freedom, and relies crucially on a reliable
description of electronic interactions. In striking contrast to these
scientific and technological needs, the actual progress in developing
techniques for describing interaction effects beyond density functional
theory \cite{kohn} in an explicitly spin-dependent manner has been slow. In
particular, one of the most important first principles many-body methods
used nowadays for describing electronic excitations, the so-called GW
method, still awaits its extension to spin-dependent interactions.

In 1965, Hedin derived a closed set of equations for the electronic Green's
function and self-energy, the screened Coulomb interaction and the
polarization of a solid \cite{hedin}. The equations provide an iterative
scheme for an expansion of the self-energy in powers of the screened
interaction. Although there is no guarantee that the iterative procedure is
absolutely convergent, as originally envisioned by Hedin, it nevertheless
provides a rigorous basis for studying the self-energy of real materials
from first-principles. In particular, the lowest order approximation leads
to the \emph{GW} approximation (GWA) \cite{hedin, ferdi, onida} which has
proven very successful in studying one-particle excitation energies of real
materials entirely from first-principles.

The original Hedin's equations were derived for a many-body Hamiltonian with
a Coulomb interaction only, without the possibility of having an explicit
spin-dependent interaction. The GWA derived from them has met tremendous
progress in describing the electronic structure and excitation spectra of
semiconductor-based systems, transition metals and oxides. Due to their
restriction to spin-independent interactions it has however not been useful
for studying correlation effects arising from spin-dependent interactions in
the classes of materials mentioned above. These interactions are crucial,
despite the tiny energy scales associated with them. To cite an example, the
conduction band spin splitting in zinc blende semiconductors arising from
spin-orbit coupling is only tens of \emph{meV} and yet it is important for
applications in spintronics since it determines the spin lifetimes as well
as inducing spin current in the absence of a magnetic field, the so-called
Rashba effect \cite{rashba}.

The aim of the present work is to present generalized Hedin's equations,
suitable for dealing with explicitly spin-dependent interactions. These
interactions may arise from relativistic effects, such as spin-orbit
coupling, or from an external perturbation like in the case of a magnetic
impurity in a semiconductor. With a spin-dependent interaction it is possible
to relate the two-particle Green's function to functional derivatives of the
one-particle Green's function with respect to an electric \emph{and} a
magnetic field. The first order term in the screened interaction of the
resulting spin-dependent Hedin's equations yields a generalization of the GW
approximation.

The Hamiltonian with a spin-dependent two-particle interaction is given by

\begin{align}
\hat{H}_{0}& =\sum_{\kappa }\int d^{3}r\ \hat{\psi}_{\kappa }^{+}(\mathbf{r}%
)h_{0}(\mathbf{r})\hat{\psi}_{\kappa }(\mathbf{r}) \\
& +\frac{1}{2}\sum_{\kappa \beta \gamma \eta }\int d^{3}rd^{3}r^{\prime }\ 
\hat{\psi}_{\kappa }^{+}(\mathbf{r})\hat{\psi}_{\beta }^{+}(\mathbf{r}%
^{\prime })v_{\kappa \gamma \beta \eta }(\mathbf{r,r}^{\prime })\hat{\psi}%
_{\eta }(\mathbf{r}^{\prime })\hat{\psi}_{\gamma }(\mathbf{r}).  \notag
\label{H}
\end{align}%
where $h_{0}$ is the one-particle Hamiltonian. The identity of the particles
implies that the second term is invariant under particle interchange: $%
v_{\kappa \gamma \beta \eta }(\mathbf{r,r}^{\prime })\leftrightarrow
v_{\beta \eta \kappa \gamma }(\mathbf{r}^{\prime }\mathbf{,r})$ \cite{fetter}%
. For this general Hamiltonian, we will %in the following 
derive the following set of generalized Hedin's equations, relating the
electronic self-energy $\Sigma $ to the Green's function $G$ and the
screened interaction $W$, using the polarization $P$ and the vertex function 
$\Lambda $:

\begin{equation}
\Sigma _{\alpha \beta }(1,2)=-\sigma _{\alpha \eta }^{I}\mathcal{G}_{\eta
\gamma }(1,4)\Lambda _{\gamma \beta }^{J}(4,2,5)W_{JI}(5,1),  \label{Sigma}
\end{equation}

\begin{equation}
W_{IJ}(1,2)=v_{IJ}(1,2)+v_{IK}(1,3)P_{KL}(3,4)W_{LJ}(4,2),  \label{W}
\end{equation}

\begin{equation}
P_{IJ}(1,2)=\sigma _{\alpha \beta }^{I}\mathcal{G}_{\beta \gamma
}(1,3)\Lambda _{\gamma \eta }^{J}(3,4,2)\mathcal{G}_{\eta \alpha }(4,1^{+}),
\label{P}
\end{equation}

\begin{eqnarray}
&&\Lambda _{\alpha \beta }^{I}(1,2,3) = \delta (1-2)\delta (2-3)\sigma
_{\alpha \beta }^{I}  \notag \\
&&+\frac{\delta \Sigma _{\alpha \beta }(1,2)}{\delta \mathcal{G}_{\gamma
\eta }(4,5)}\mathcal{G}_{\eta \eta ^{\prime }}\mathcal{(}4,6)\Lambda _{\eta
^{\prime }\kappa }^{I}(6,7,3)\mathcal{G}_{\kappa \gamma }(7,5).\mathcal{%
7777777}  \label{vertex}
\end{eqnarray}

Here, $\sigma ^{i}$, i=x,y,z, are the Pauli spin matrices and $\sigma ^{0}$ is defined
to be a 2x2 unit matrix. Capital letter indices run over 0,x,y,z, while
Greek letters take the values $\pm 1$. We use the following common shorthand
notation: $(\mathbf{x}\tau )$ is represented by a number, repeated indices
are summed and repeated variables represented by numbers are integrated,
unless they appear on both sides of the equation. For clarity, we further
adopt a notation that quantities with subscript denoted by capital letter do
not depend on spin. The spin-dependent interaction has been expanded in the
Pauli and unit matrices as

\begin{equation}
v_{\alpha \eta \kappa \gamma }(1,2)=\sigma _{\alpha \eta
}^{I}v_{IJ}(1,2)\sigma _{\kappa \gamma }^{J}.  \label{v}
\end{equation}%

The spin-Hedin equations can e.g.~be applied to Hamiltonians containing
interactions of the following form.

\begin{equation}
v_{\alpha \gamma \beta \eta }(\mathbf{r,r}^{\prime })=\left\{ 
\begin{array}{c}
\sigma _{\alpha \gamma }^{0}\sigma _{\beta \eta }^{0}/|\mathbf{r-r}^{\prime
}|, \\ 
\sigma _{\alpha \gamma }^{i}J_{ij}(\mathbf{r,r}^{\prime })\sigma _{\beta
\eta }^{j}, \\ 
\sigma _{\alpha \gamma }^{i}\mu _{i}(\mathbf{r,r}^{\prime })\sigma _{\beta
\eta }^{0}%
\end{array}%
\right.  \label{interaction}
\end{equation}%
where the first is the usual Coulomb interaction, the second a spin-spin
interaction and the third a spin-orbit interaction, which contains the
angular momentum operator.

The general structure of the set
of equations (\ref{Sigma}) -- (\ref{vertex})
is remarkably close to the usual
Hedin's equations. However, the self-energy now depends on the spin
variable, and the screened interaction $W$ as well as the polarization
function $P$ acquire a matrix form expressing an interplay between the
charge and spin channels. Thus, a polarization function $P_{0i}$, for
example, describes a charge density response of the system with respect to a
perturbing magnetic field in the \emph{i}-direction. When $P$ is used in (%
\ref{W}) the charge channel of the screened interaction experiences the
effects of the spin interactions and vice versa.

Most importantly, one can construct from these equations a spin-dependent
generalization of Hedin's GWA, by approximating the vertex functions by

\begin{equation}
\Lambda _{\alpha \beta }^{I}(1,2,3)=\delta (1-2)\delta (2-3)\sigma _{\alpha
\beta }^{I}.  \label{vertex0}
\end{equation}%
The polarization then becomes

\begin{equation}
P_{IJ}(1,2)=\sigma _{\alpha \beta }^{I}\mathcal{G}_{\beta \gamma
}(1,2)\sigma _{\gamma \eta }^{J}\mathcal{G}_{\eta \alpha }(2,1^{+})
\label{P0}
\end{equation}%
yielding the self-energy

\begin{equation}
\Sigma _{\alpha \beta }^{GW}(1,2)=-\sigma _{\alpha \eta }^{I}\mathcal{G}%
_{\eta \gamma }(1,2)\sigma _{\gamma \beta }^{J}W_{JI}(2,1).  \label{SigmaGW}
\end{equation}%
Before proceeding further, let us interpret the meaning of the resulting
polarization and the spin-dependent self-energy. Consider first the case $%
I=J=0$ (charge channel) giving $P_{00}(1,2)=\mathcal{G}_{\alpha \gamma }(1,2)%
\mathcal{G}_{\gamma \alpha }(2,1^{+})$ and $\Sigma _{\alpha \beta }(1,2)=-%
\mathcal{G}_{\alpha \beta }(1,2)W_{00}(2,1)$. If the Green function is
diagonal in spin space we recover the polarization and the self-energy in
the original Hedin's equations. However, for a system with an existing spin
structure, such as a non-colinear spin, the Green function possesses
non-diagonal spin components. This case is a generalization of the original
Hedin's equations to spin-dependent Green's function and self-energy with
purely Coulombic interaction. It emerges naturally in the present
formulation as a special case where spin interactions are absent. 

Let us now consider the case when the interaction is spin dependent, which
may arise from purely spin-spin interaction or spin-orbit coupling, among
other possibilities. The exchange-correlation effects on the Green function
of up spin is illustrated in Fig. 1 (lower panel). A particle of up spin $%
\mathcal{G}_{\uparrow \uparrow }^{0}$ enters the self-energy $\Sigma
_{\uparrow \uparrow }$. Upon entering the self-energy the electron spin is
flipped to down spin by a spin operator $\sigma _{\uparrow \downarrow }^{i}$%
and a magnon represented by $W_{ij}$ is emitted. Upon leaving the
self-energy the spin operator $\sigma _{\downarrow \uparrow }^{j}$ causes
the electron to reabsorb the magnon and return to its original up spin
configuration. This process is analogous to the original Hedin's GWA whereby
an electron emits and absorbs a plasmon but without the possibility of spin
flip (upper panel).

\begin{figure}[th]
\begin{center}
\includegraphics[width=0.5\textwidth]{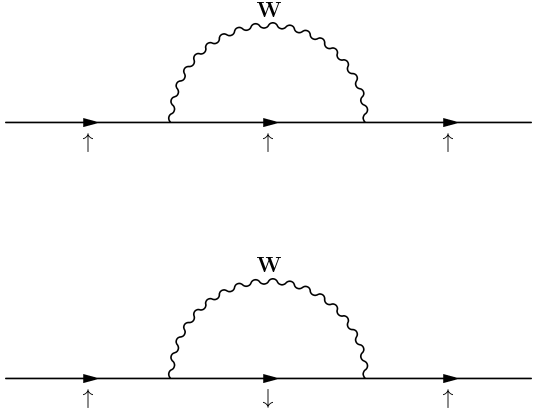}
\end{center}
\caption{Diagrams for the self-energy within the spin-dependent GWA,
compared with the conventional GWA see text}
\end{figure}

The derivation of the generalized Hedin's equations closely follows Hedin's
original work, using Schwinger's functional derivative technique. Since many
problems related to spin degrees of freedom involves temperature, we work in
the finite-temperature formalism but of course the zero-temperature version
readily follows. Using the Heisenberg equations of motion we obtain

\begin{align}
& \left[ \frac{\partial }{\partial \tau }+h_{0}(\mathbf{x})-\mu \right] 
\mathcal{G}_{\alpha \beta }(\mathbf{x}\tau ,\mathbf{x}^{\prime }\tau
^{\prime })  \notag \\
& +\int d^{3}r\ v_{\kappa \gamma \alpha \eta }(\mathbf{r,x})\mathcal{G}%
_{\eta \beta \gamma \kappa }^{(2)}(\mathbf{x}\tau ,\mathbf{x}^{\prime }\tau
^{\prime },\mathbf{r}\tau ,\mathbf{r}\tau ^{+})  \notag \\
& =-\delta _{\alpha \beta }\delta (\mathbf{x-x}^{\prime })\delta (\tau -\tau
^{\prime }).  \label{eqnmotion}
\end{align}%
To utilize the Schwinger functional derivative technique we work in the
Dirac or interacting representation and define the Green functions as
follows.

\begin{equation}
\mathcal{G}_{\alpha \beta }(1,2)=-\frac{\left\langle \mathcal{T}\ [\mathcal{%
\hat{S}}\hat{\psi}_{\alpha }(1)\hat{\psi}_{\beta }^{+}(2)]\right\rangle }{%
\left\langle \mathcal{\hat{S}}\right\rangle }  \label{G1}
\end{equation}

\begin{equation}
\mathcal{G}_{\alpha \beta \eta \gamma }^{(2)}(1,2,3,4)=\frac{\left\langle 
\mathcal{T}\ [\mathcal{\hat{S}}\hat{\psi}_{\alpha }(1)\hat{\psi}_{\eta }(3)%
\hat{\psi}_{\gamma }^{+}(4)\hat{\psi}_{\beta }^{+}(2)]\right\rangle }{%
\left\langle \mathcal{\hat{S}}\right\rangle }  \label{G2}
\end{equation}%
where $\mathcal{T}$ is the imaginary-time-ordering operator and%
\begin{equation}
\mathcal{\hat{S}}=\mathcal{T}\ \exp \left[ -\int_{0}^{\beta }d\tau ^{ }\ 
\hat{\phi}(\tau)\right] .  \label{S}
\end{equation}

In the original Hedin's derivation a probing electric field was applied to
obtain an equation relating the two-particle Green's function to the
functional derivative of the one-particle Green's function with respect to
the applied electric field. Since we have spin-dependent interactions we
have to consider not only a probing electric field, $\varphi _{0}(\mathbf{r}%
\tau )$, but also a probing magnetic field, $\varphi _{i}(\mathbf{r}\tau ),\
i=x,y,z$. The electric and magnetic fields are given by

\begin{equation}
\hat{\phi}(\tau )=\int d^{3}r\ \varphi _{I}(\mathbf{r}\tau )\hat{\psi}%
_{\alpha }^{+}(\mathbf{r}\tau )\sigma _{\alpha \beta }^{I}\hat{\psi}_{\beta
}(\mathbf{r}\tau ).  \label{phi}
\end{equation}

Noting that $\delta \mathcal{\hat{S}}/\delta \varphi _{I}(1)=-\mathcal{T[}%
\mathcal{\hat{S}\hat{\sigma}}^{I}(1)]$ it can be shown that \cite{ferdispin}%
\begin{eqnarray}
&&\frac{\delta \mathcal{G}_{\eta \beta }(1,2)}{\delta \varphi _{I}(3)}
\label{dGdphi} \\
&=&\left[ \mathcal{G}_{\eta \beta }(1,2)\mathcal{G}_{\gamma \kappa
}(3,3^{+})-\mathcal{G}_{\eta \beta \gamma \kappa }^{(2)}(1,2,3,3^{+})\right]
\sigma _{\kappa \gamma }^{I}.
\end{eqnarray}%
This important relation allows us to replace the two-particle Green's
function by a functional derivative of the one-particle Green's function
with respect to the applied electric and magnetic field. Using the above
relation we define the mass operator $\mathcal{M}$ as follows.

\begin{eqnarray}
&&\mathcal{M}_{\alpha \gamma }(1,3)\mathcal{G}_{\gamma \beta }(3,2)  \notag
\\
&=&v_{\kappa \gamma \alpha \eta }(3,1)\mathcal{G}_{\eta \beta \gamma \kappa
}^{(2)}(1,2,3,3^{+})  \notag \\
&=&v_{IJ}(3,1)\sigma _{\alpha \eta }^{J}\left[ \mathcal{G}_{\eta \beta
}(1,2)\rho _{I}(3)-\frac{\delta \mathcal{G}_{\eta \beta }(1,2)}{\delta
\varphi _{I}(3)}\right] .  \label{M}
\end{eqnarray}%
The charge and spin density is given by

\begin{equation}
\rho _{I}(1)=\mathcal{G}_{\gamma \kappa }(1,1^{+})\sigma _{\kappa \gamma
}^{I}.  \label{density}
\end{equation}%
Using the identity

\begin{equation}
(\delta \mathcal{G}/\delta \varphi )=-\mathcal{G}(\delta \mathcal{G}%
^{-1}/\delta \varphi )\mathcal{G},  \label{identity}
\end{equation}%
multiplying by $\mathcal{G}^{-1}$ on the right and taking out the
Hartree-like term (the first term) we obtain from (\ref{M}) an exact
expression for the self-energy:

\begin{equation}
\Sigma _{\alpha \beta }(1,2)=v_{IJ}(3,1)\sigma _{\alpha \gamma }^{J}\mathcal{%
G}_{\gamma \eta }(1,4)\frac{\delta \mathcal{G}_{\eta \beta }^{-1}(4,2)}{%
\delta \varphi _{I}(3)}.  \label{Sigmaexact}
\end{equation}%
As pointed out in \cite{ferdispin} in the absence of an explicit
two-particle spin interaction, only $I=J=0$ components survive and the
self-energy depends only on variation with respect to the electric field $%
\varphi _{0}$. Application of a magnetic field introduces a one-particle
term in the Hamiltonian but it does not affect the self-energy.

Next, we identify the generalized Hartree potential

\begin{equation}
V_{I}^{H}(1)=v_{IJ}(1,3)\rho _{J}(3)  \label{Hartree}
\end{equation}%
and the total field $\Phi _{I}=\varphi _{I}+V_{I}^{H}$. The vertices can
then be defined as follows:

\begin{equation}
\Lambda _{\alpha \beta }^{I}(1,2,3)=-\frac{\delta \mathcal{G}_{\alpha \beta
}^{-1}(1,2)}{\delta \Phi _{I}(3)}.  \label{vertexdef}
\end{equation}%
As usual, the dielectric function is related to the derivative of the total
field with respect to the applied one 
\begin{equation}
\frac{\delta \Phi _{I}(1)}{\delta \varphi _{J}(2)}=\varepsilon
_{IJ}^{-1}(1,2)=\delta (1-2)\delta _{IJ}+\frac{\delta V_{I}^{H}(1)}{\delta
\varphi _{J}(2)},  \label{diel}
\end{equation}%
Thus, using the chain rule

\begin{equation}
-\frac{\delta \mathcal{G}_{\alpha \beta }^{-1}(1,2)}{\delta \varphi _{I}(3)}%
=\Lambda _{\alpha \beta }^{J}(1,2,4)\varepsilon _{JI}^{-1}(4,3),
\end{equation}%
and inserting this equation into (\ref{Sigmaexact}) we obtain Eq. (\ref%
{Sigma}) 
with $W_{IJ}(1,2)=\epsilon _{IK}^{-1}(1,3)v_{KJ}(3,2)$. The charge and spin
vertex equations are given by

\begin{align}
\Lambda _{\alpha \beta }^{I}(1,2,3)& =\delta (1-2)\delta (2-3)\sigma
_{\alpha \beta }^{I}+\frac{\delta \Sigma _{\alpha \beta }(1,2)}{\delta \Phi
_{I}(3)}  \notag \\
& =\delta (1-2)\delta (2-3)\sigma _{\alpha \beta }^{I}  \notag \\
& +\frac{\delta \Sigma _{\alpha \beta }(1,2)}{\delta \mathcal{G}_{\gamma
\eta }(4,5)}\mathcal{G}_{\eta \eta ^{\prime }}\mathcal{(}4,6)\Lambda _{\eta
^{\prime }\kappa }^{I}(6,7,3)\mathcal{G}_{\kappa \gamma }(7,5)\mathcal{%
7777777}
\end{align}%
where we have made use of the chain rule, the identity (\ref{identity}),
and the definition of the vertex in (\ref{vertexdef}).

We now establish the equation for the screened interaction. The polarization
is given by the variation of the density with respect to the total field.
Using (\ref{density}), (\ref{identity}), and (\ref{vertexdef})

\begin{align}
P_{IJ}(1,2)& =\frac{\delta \rho _{I}(1)}{\delta \Phi _{J}(2)}  \notag \\
& =\sigma _{\alpha \beta }^{I}\frac{\delta \mathcal{G}_{\beta \alpha
}(1,1^{+})}{\delta \Phi _{J}(2)}  \notag \\
& =-\sigma _{\alpha \beta }^{I}\mathcal{G}_{\beta \gamma }(1,3)\frac{\delta 
\mathcal{G}_{\gamma \eta }^{-1}(3,4)}{\delta \Phi _{J}(2)}\mathcal{G}_{\eta
\alpha }(4,1^{+})  \notag \\
& =\sigma _{\alpha \beta }^{I}\mathcal{G}_{\beta \gamma }(1,3)\Lambda
_{\gamma \eta }^{J}(3,4,2)\mathcal{G}_{\eta \alpha }(4,1^{+}).
\end{align}%
Expressing the dielectric function (\ref{diel}) as 
\begin{align}
\varepsilon _{IJ}^{-1}(1,2)& =\delta (1-2)\delta _{IJ}+\frac{\delta
V_{I}^{H}(1)}{\delta \rho _{M}(3)}\frac{\delta \rho _{M}(3)}{\delta \Phi
_{K}(4)}\frac{\delta \Phi _{K}(4)}{\delta \varphi _{J}(2)}  \notag \\
& =\delta (1-2)\delta _{IJ}+v_{IM}(1,3)P_{MK}(3,4)\varepsilon _{KJ}^{-1}(4,2)
\label{dieleqn}
\end{align}%
and multiplying by the bare interaction, we obtain Eq. (\ref{W}). 
The response function, $R=\delta \rho /\delta \varphi $, satisfies a similar
equation:

\begin{equation}
R_{IJ}(1,2)=P_{IJ}(1,2)+P_{IK}(1,3)v_{KL}(3,4)R_{KJ}(4,2)  \label{R}
\end{equation}%
which follows from $\delta \rho /\delta \varphi =\delta \rho /\delta \Phi
\times \delta \Phi /\delta \varphi =P\varepsilon ^{-1}$. The above three
equations for $\varepsilon ^{-1}$, $W$, and $R$ are equivalent. This
concludes our derivation of the spin-dependent Hedin's equations.

In the following we first comment on some practical aspects of the
generalized GWA; we then conclude by discussing possible applications and
further developments.

We note that while the magnetic Hedin's equations are exact, 
the solution depends crucially on the starting Green's function, when an
iterative scheme is used. % to solve them. 
For example, if we consider an Ising interaction and choose a starting Green's
function corresponding to the mean-field solution, solving the magnetic
Hedin's equations iteratively will not generate other spin configurations
because there is no spin-flip term in the Ising model. More explicitly, the
polarization in (\ref{P0}) with the mean-field Green's function will only
possess the z-component $P_{zz}$, which in turns will only generate $W_{zz}$%
. Thus, the self-energy in (\ref{SigmaGW}) will not have off-diagonal
components in spin space that would allow for spin-flip processes yielding
spin configurations other than the mean-field one. To solve this problem,
one must start with a starting Green's function that corresponds to a multi
configuration. This problem does not arise in the general spin spiral case,
where the spin at each point may have different magnitude and direction. 
The non-interacting Green's function \cite{footnote}
\begin{equation}
\mathcal{G}_{\alpha \beta }^{0}(\mathbf{r,r}^{\prime };i\omega )=\sum_{n}%
\frac{\phi _{n\alpha }^{\ast }(\mathbf{r})\phi _{n\beta }(\mathbf{r}^{\prime
})}{i\omega -\varepsilon _{n}}  \label{GLDA}
\end{equation}%
possesses non-diagonal components in the spin space which contribute to the
charge channel of the polarization.

\begin{align}
P_{00}(1,2)& =\sigma _{\alpha \beta }^{0}\mathcal{G}_{\beta \gamma
}(1,2)\sigma _{\gamma \eta }^{0}\mathcal{G}_{\eta \alpha }(2,1^{+})
\label{P00} \\
& =\mathcal{G}_{\alpha \gamma }(1,2)\mathcal{G}_{\gamma \alpha }(2,1^{+})
\notag
\end{align}%
and the self-energy naturally acquires non-diagonal components in the spin
space.

\begin{equation}
\Sigma _{\alpha \beta }^{GW}(1,2)=-\mathcal{G}_{\alpha \beta
}(1,2)W_{00}(2,1).  \label{SigmaGW0}
\end{equation}%
Using the self-energy one solves the quasiparticle equation%
\begin{equation}
\left( 
\begin{array}{ll}
h^{0}+\Sigma _{\uparrow \uparrow }^{GW}(E_{n}) & \ \ \ \ \Sigma _{\uparrow
\downarrow }^{GW}(E_{n}) \\ 
\ \ \ \ \Sigma _{\downarrow \uparrow }^{GW}(E_{n}) & h^{0}+\Sigma
_{\downarrow \downarrow }^{GW}(E_{n})%
\end{array}%
\right) \left( 
\begin{array}{l}
\psi _{n\uparrow } \\ 
\psi _{n\downarrow }%
\end{array}%
\right) =E_{n}\left( 
\begin{array}{l}
\psi _{n\uparrow } \\ 
\psi _{n\downarrow }%
\end{array}%
\right)  \label{QPspin}
\end{equation}%
which contains the effects of the non-diagonal spin components of the
self-energy.

A non-interacting Green's function constructed from a Hamiltonian containing
a spin-orbit interaction will also have non-diagonal spin components. A
similar formulation as above can be applied to this situation. The screened
interaction will contain charge-spin coupling components and the self-energy
will take a more general form as given in (\ref{SigmaGW}). Correlation
effects on orbital moments and spin densities can then be accessed including
life-time effects.

In conclusion, we have derived the spin-dependent generalization of the
original set of Hedin's equations. Their first order term (in the
spin-dependent screened interaction) leads to a spin-dependent GW
approximation. These equations allow for a truly first-principles study of a
wide range of problems where correlation effects induced by spin
interactions play a crucial role in determining physical properties.
Applications to nanoscale magnetic systems ranging from quantum dots,
quantum wires to impurities or nanoparticles, as well as to films, surface
and interface problems now come into reach \cite{hernando, bode}. Also,
developments at the interface of many-body perturbation theory and
time-dependent density functional theory as proposed in \cite{bruneval,
marini} are readily generalized e.g.~to relativistic interactions. The same
is true for combined GW and dynamical mean field techniques \cite{gwdmft}.


\begin{thebibliography}{99}
\bibitem{footnote} We refer to the one-particle part of the problem as
non-interacting, even if it will in general contain interactions on a mean
field or density functional \cite{kohn} level, giving rise to the
effective (mean field or Kohn Sham) potential $V^{xc}$.
\end{thebibliography}
\end{document}